# Thresholds in choice behaviour and the size of travel time savings

Andy Obermeyer,[a] Martin Treiber,[a] Christos Evangelinos [b]

[a] Institute for Transport and Economics, Friedrich List Faculty of Transportation and Traffic Sciences, Technische Universität Dresden (Dresden University of Technology), Germany.

[b] School of Business and Management, International University of Applied Sciences Bad Honnef/Bonn, Germany.

*E-mail:* andy.obermeyer@tu-dresden.de, treiber@vwi-dresden.de, c.evangelinos@iubh.de

**Abstract**

Travel time savings are usually the most substantial economic benefit of transport infrastructure projects. However, questions surround whether small time savings are as valuable per unit as larger savings. Thresholds in individual choice behaviour are one reason cited for a discounted unit value for small time savings. We demonstrate different approaches for modelling these thresholds using synthetic and stated choice data. We show that the consideration of thresholds is important, even if the discounted unit value for small travel time savings is rejected for transport project appraisal. If an existing threshold is ignored in model estimation, the value of travel time savings will be biased. The presented procedure might also be useful to model thresholds in other contexts of choice behaviour.

# 1. Introduction

One of the main outcomes of transport infrastructure improvements are travel time reductions. Their evaluation therefore plays a major role in infrastructure planning and assessment. For instance, travel time reductions may change individual route or mode choices and, as a consequence, may affect the flow of traffic. Usually, benefit-cost analyses are performed in order to assess whether a project is beneficial for society or not. In such analyses, travel time savings usually comprise a substantial economic benefit of transport infrastructure projects. Welch and Williams (1997, p. 231), for example, find that time-travel savings represent between 70 to 90 per cent of total benefits. Others as Fosgerau and Jensen (2003), Mackie et al. (2001, p. 91), and Hensher (2001, p. 71) report shares of 80, 80, and 60 per cent, respectively. Nowadays, however, other important components are considered as well, as for example, the reliability of travel time (Small, 2012, p. 2). For a typical urban automobile commuter trip in the US, Small and Verhoef (2007, p. 98) show that travel time accounts for 34 and reliability for 11 per cent of the short-run social variable costs.[1] Nevertheless, the value of travel time savings is still a key element in transport project appraisal. Furthermore, Welch and Williams (1997, p. 233) state that these benefits are mainly caused by small time savings (on the order of a few minutes). Metz (2008a) criticizes the current practice of project appraisal based on travel time savings and argues that at least in the long run providing access to destinations is the main purpose of transport. Several authors, however, disagree with Metz's (2008a) conclusions and argue that the valuation of travel time savings is an appropriate approach for benefit-cost analyses (Ironmonger and Norman, 2008; Mackie, 2008; Van Wee and Rietveld, 2008).[2]

There has been a long and ongoing debate on how to treat small travel time savings, since thresholds in individual choice behaviour might be present for small travel time savings. The discussion usually focuses on the monetization of travel time savings.[3] In several studies it has been found that small travel time savings are less valued by travellers on a unit basis than larger ones (e.g. Bates and Whelan, 2001; Fosgerau, 2007; Gunn, 2001; Hultkrantz and Mortazavi, 2001; Mackie et al., 2003).[4] As demonstrated by Welch and Williams (1997) in a case study, the application of discounted unit values for small time savings can have a significant influence on overall project benefits. One of the main arguments in favour of a reduced monetary value for small travel time savings is that people cannot make effective use of them. However, counterarguments hold that, in the long run, small savings aggregate, such that they can be effectively utilized (e.g. Fowkes, 1999; Mackie et al., 2001). Alternatively, small time savings may be rejected by individuals because the cognitive decision costs of evaluating the alternatives might exceed the possible benefit that could be gained (Hultkrantz and Mortazavi, 2001, p. 290). For the matter of project appraisal it is relevant, whether a threshold is a consequence of real social costs or just a behavioural element people show in stated choice studies (Hultkrantz and Mortazavi, 2001, p. 294). In addition, the fact that travellers might not recognize small time differences (because they are below their cognition threshold) does not mean the benefits associated with them are lost (Mackie et al., 2001). Following these arguments, small travel time savings should be valued the same on a unit basis as large ones in benefit-cost analyses. However, for model estimation, the presence of thresholds in individual choice behaviour might still be important.

---

[1] For a brief overview on the value of reliability see Small and Verhoef (2007, p. 49-55).

[2] The statements of Metz (2008a) are also discussed by Givoni (2008), Lyons (2008), Noland (2008), Schwanen (2008), and Metz (2008b).

[3] Another important issue is the potential error in measurement of small travel time savings in transport models. However, this paper deals with another subject.

[4] Interestingly, the authors (Bates and Whelan, 2001; Mackie et al., 2003) question their empirical results and argue for a constant unit value of travel time savings instead.

In this paper, we focus on empirical issues in estimating thresholds with respect to travel time differences by discrete choice models and the consequences of ignoring them in model estimation. From our point of view, this issue has to be addressed separately from the question whether thresholds should be considered in benefit-cost analyses. As we show, estimated asymptotic values of travel time savings can differ substantially from that of a model ignoring these thresholds. Furthermore, the consideration of thresholds may be important for predicting choice behaviour. Two new functions will be presented that permit the modelling of smooth thresholds. These functions can easily be applied in any estimation tool for discrete choice analysis that can handle non-linear utility functions. We demonstrate the usefulness of these functions with synthetic data and apply them to stated choice data. It is not the scope of this paper to answer the question whether thresholds should be considered in benefit-cost analyses or not. However, as this issue is inevitably connected to the empirical analysis presented in the following, the most common arguments that appear in the literature regarding this question will be summarized.

The explicit consideration of thresholds with respect to travel time savings is a rather underrepresented topic in the literature on travel choice modelling. Essentially, such approaches focus either on utility or attributes. The former, which is based on indifference (utility) thresholds, has been modelled, for example, by Krishnan (1977), Lioukas (1984), and Cantillo et al. (2010). We, however, will concentrate on attribute thresholds. Work in this area has been done, for example, by Cantillo et al. (2006) and Li and Hultkrantz (2004). Empirical results support the existence of utility as well as attribute thresholds.

One may notice that prospect theory posits increased rather than decreased sensitivity near a reference point. However, this fact can also explain a lower value of time for small time differences as shown by Hjorth and Fosgerau (2012). They applied a sophisticated power function transformation to time and cost differences to model the main properties of prospect theory's value function – namely, increased sensitivity for small differences from the reference and loss aversion.[5] An increasing value of time with the size of time difference is explained by a relatively stronger diminishing sensitivity for money than for time.

The structure of the paper is as follows. In section 2 and 3, the modelling approach is described and tested using synthetic data. Section 4 presents the calculation of the value of travel time savings and discusses the topic of project scheme appraisal and time thresholds. In section 5, the modelling approach is applied to stated choice data. Finally, section 6 concludes with a discussion.

## 2. Modelling Approach

We model the choice between two alternatives of the same category that are characterised by the same attributes, i.e., unlabelled alternatives (e.g. route choice between a cheap but slow and a fast but expensive train connection). The modelling approach focuses on detection of possible deviations in the sensitivity to attribute differences between both alternatives, if these differences are small. The aim is to test whether travellers exhibit different sensitivities between large and small travel time differences. In this case the rate of substitution between travel cost and travel time will be different between small and large changes in travel time (assuming constant cost sensitivity).

It is assumed that trip makers always choose the option with the highest utility, which is decomposed into a deterministic ($V$) and a stochastic ($\varepsilon$) part. The stochastic component is assumed to be iid Gumbel, and, therefore, the difference between the two stochastic components is logistically distributed. In the following, we consider just the utility difference between the two alternatives, because this is what matters for the choice decision. The utility difference is a function of the attribute differences. To model potentially different sensitivities depending on the size of time differences, an attribute transformation function is applied. The parameter $\alpha_r$ of the transformation function has to be estimated along with the remaining coefficients of the model.

We assume that the utility difference $\Delta U$ is separable in time differences between alternatives $\Delta T$ (in minutes) and cost differences $\Delta C$ (in CHF) according to

$$\Delta U(\Delta T, \Delta C) = \Delta V(\Delta T, \Delta C) + \Delta \varepsilon = \beta_T * f_r(\Delta T, \alpha_r) + \beta_C \Delta C + \Delta \varepsilon \qquad (1)$$

---

[5] For an overview on the topic of reference point formation and loss aversion in the framework of choice modelling see Stathopoulos and Hess (2012).

and that the time component is non-linear according to the following three specifications of the transformation function $f_r(\Delta T, \alpha_r)$ (cf. Figure 1).[6]

$$f_{HTF}(\Delta T, \alpha_{HTF}) = \begin{cases} 0 & \text{abs}(\Delta T) < \alpha_{HTF} \\ \text{sign}(\Delta T) * (\text{abs}(\Delta T) - \alpha_{HTF}) & \text{abs}(\Delta T) \geq \alpha_{HTF} \end{cases} \quad (2)$$

$$f_{STF1}(\Delta T, \alpha_{STF1}) = \Delta T - \alpha_{STF1} \tanh\left(\frac{\Delta T}{\alpha_{STF1}}\right) \quad (3)$$

$$f_{STF2}(\Delta T, \alpha_{STF2}) = \Delta T \left(1 - 1/\sqrt{\left(\frac{\Delta T}{\alpha_{STF2}}\right)^2 + 1}\right) \quad (4)$$

Eq. ( 2 ) is a hard threshold function (HTF), which is usually applied to model thresholds. It is a piecewise linear function, where the slope within the threshold area is zero. For functions ( 3 ) and ( 4 ), called soft threshold functions (STF), the slope is continuously increasing from zero to the limit of one. Figure 1 depicts the three different transformation functions for a threshold of five minutes.

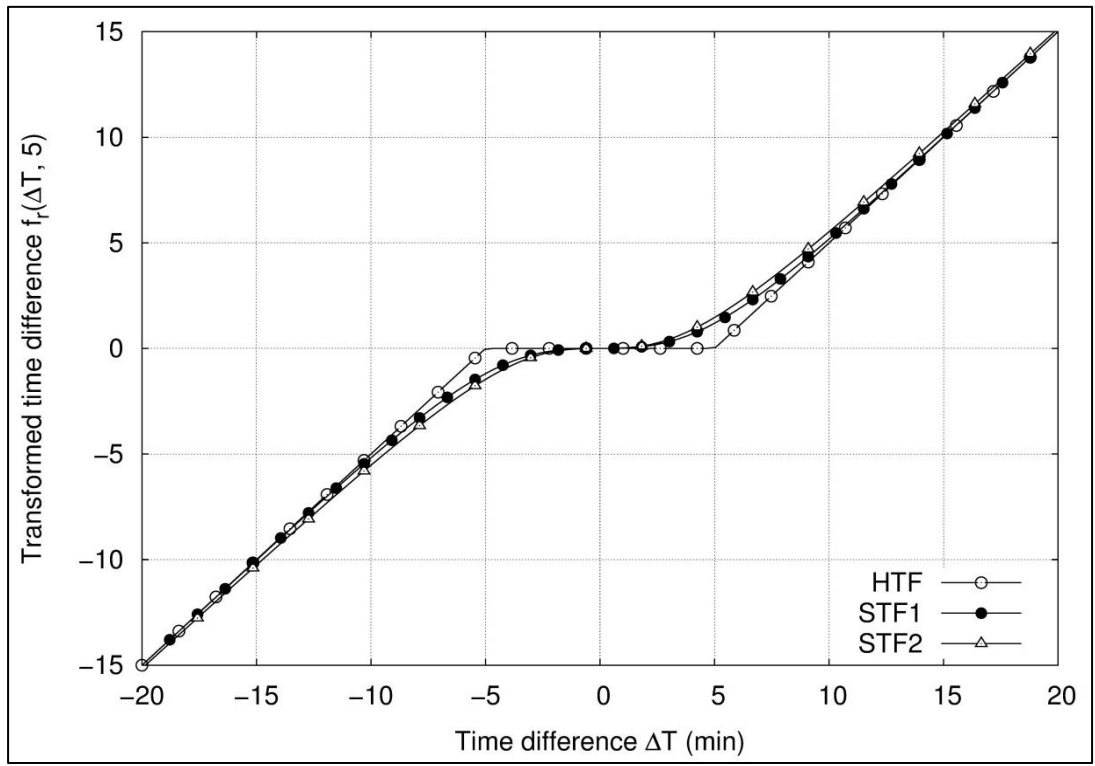

**Figure 1:** Generic form of the transformation functions ( 2 ) - ( 4 ); for illustration purposes these functions are plotted for a threshold parameter $\alpha_r = 5$ (min).

The HTF models the extreme case of no sensitivity at all within the threshold. This is the common understanding of a threshold. However, in the following, the term threshold will be used to describe the area with significantly reduced sensitivity of the STF as well. Both STF are approximations to the HTF. The limit of the slopes of the STF for large positive and negative time differences is one and hence the asymptotes of the STF correspond to the HTF (if $\alpha_{HTF} = \alpha_{STF1} = \alpha_{STF2}$). The smooth functions can easily be implemented in any estimation tool that can handle non-linear utility functions, which is not necessarily the case for the HTF.

---
[6] For procedures to estimate such models see, for example, Train (2009).

Another option to model reduced sensitivity for small time differences is to employ a power function transformation.[7]

$$f_{Power}(\Delta T, \alpha_{Power}) = sign(\Delta T) * abs(\Delta T)^{\alpha_{Power}} \qquad (5)$$

Usually, this kind of transformation is used to model increased sensitivity around a reference point as predicted by prospect theory.[8] An exponent greater than one means a reduced sensitivity for small differences. However, in contrast to the STF, the power function exhibits no limit for the slope when differences tend to infinity. This might be problematic for obtaining correct estimates, as we will show with synthetic data below.

## 3. Application to synthetic data

To test the different specifications regarding their goodness of fit, a synthetic database with two alternatives has been set up. For 5 000 database records, time and cost differences as well as a logistically distributed error component for the differences (corresponding to a logit model) have been generated. Cost and time differences have been assumed to be independent and uniformly distributed in the range of [-10 CHF, 10 CHF] and [-25 min, 25 min], respectively. Dominant choice sets (with only positive or negative time and cost differences) have been excluded. The deterministic utility differences have been calculated according to the hard threshold function. In line with the general procedure in discrete choice modelling, the alternative with the greatest total utility (i.e., positive difference to the other alternative) has been selected. Based on this data, estimations have been carried out with all four presented transformations.[9] Table 1 summarises the parameters used in the data generation process and the estimated values. Plots of $\Delta V$ against $\Delta T$ for all functions can be found in Figure 2.

---

[7] Commonly, power function transformations are somewhat more sophisticated, allowing different sensitivities for gains and losses (e.g. Hjorth and Fosgerau, 2012). However, for the kind of data used here, this is not necessary.

[8] Both STF can be adjusted to incorporate increased sensitivity for small attribute differences by including a further parameter. In ( 3 ) $\alpha_{STF1}$ in front of the hyperbolic tangent and in ( 4 ) the numerator one have to be replaced by a separate parameter. This is useful for testing increased cost sensitivity, which has a similar effect on the value of time as a time threshold. However, in this paper we concentrate on time thresholds.

[9] All estimations have been carried out with Python Biogeme.

**Table 1:**
Estimation results for synthetic data.

| Variable | Synthetic | Linear | HTF | STF1 | STF2 | Power |
|---|---|---|---|---|---|---|
| Cost | -0.600 | -0.630 * | -0.596 * | -0.598 * | -0.598 * | -0.602 * |
|  |  | (0.12) | (0.83) | (0.92) | (0.92) | (0.92) |
| Time | -0.100 | -0.080 * | -0.106 * | -0.113 * | -0.119 * | -0.013 + |
|  |  | (0.00) | (0.47) | (0.35) | (0.26) | (0.00) |
| Alpha | 5.000 | --- | 5.410 * | 6.34 * | 7.48 * | 1.600 * |
|  |  |  | (0.69) | (0.45) | (0.29) | [0.00] |
| VTTS [a] | 10.00 | 7.62 | 10.67 | 11.33 | 11.94 | --- |
| Null-LL |  |  |  | -3465.736 |  |  |
| Final-LL | --- | -1787.714 | -1779.042 | -1779.105 | -1779.051 | -1779.064 |

\*, #, + Significant on 1%, 5%, and 10% levels, respectively.
(.) p-value for null hypothesis that parameter is equal to its target value (synthetic column).
[.] p-value for null hypothesis that parameter is equal to one.
[a] Asymptotic value of travel time savings in CHF per hour. See section 4.

The HTF and the two STF apparently fit really well and reproduce the target values.[10] Not surprisingly, the threshold width of the HTF is closest to its original since this function determines the data generation process. The threshold parameters of the STF show a somewhat greater difference to the predefined one. However, since the STF are smooth approximations to the HTF, this deviation is not surprising and, moreover, not significant. Despite the good fit of the power function, the estimated time coefficient is significantly different from its target value. This is a consequence of the infinite slope of the power function for large attribute levels. The power is significantly larger than one, indicating a reduced sensitivity for small time differences. However, to avoid a steeply increasing slope, the time sensitivity coefficient needs to be small. Considering that the data is generated under the assumption of a hard threshold, the procedure above points out which function allows for correct parameter estimation if the extreme case of no sensitivity at all within the threshold area is present in peoples' behaviour. Strictly speaking, it has been just shown that the power function performs worse if and only if a hard threshold is present in the data. It is easy comprehensible that the STF will perform relatively better if the threshold in the data is not hard but soft. The power function, however, provides an incorrect estimate of the time coefficient as long as people do not exhibit an infinite sensitivity for large differences, although the overall model fit might be good. A purely linear specification has also been estimated to examine the error when ignoring the threshold. Although the p-value has fallen dramatically, the cost coefficient is still not significantly different from its target. However, the time coefficient has not been reproduced correctly. In general, we observe that many observations are necessary to detect an existing threshold. For 5 000 observations the log-likelihood difference between the linear and the threshold models is just about 9 units,[11] which means an average improvement of roughly 1 log-likelihood unit per 500 observations in this situation.

---

[10] Estimated coefficients are not significantly different from target values (cf. Table 1).

[11] Nonetheless, the threshold models are significantly better than the linear model.

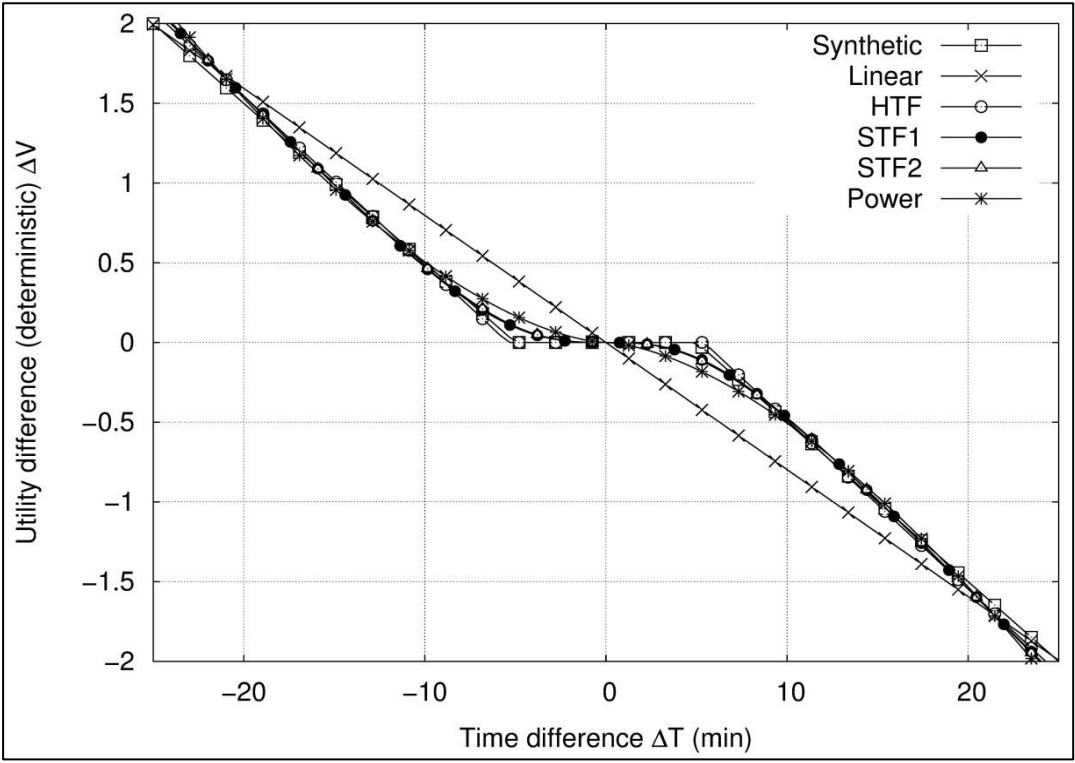

**Figure 2:** Estimated utility functions for the indicated transformation functions for synthetic data (open squares) with 5 000 observations.

To examine the validity of using the results of just one simulation instance for further calculations we generated 500 databases with 5 000 simulated choices each. Estimations have been carried out for each of the five functions shown above, resulting in 500 estimated parameter sets for each function. Based on these estimates the mean and the standard error have been calculated for each coefficient and function. Results are summarised in Table 2. For comparison, the corresponding standard errors of the single simulation instance reported in Table 1 are included in Table 2. Estimated means and standard errors based on the 500 simulation runs are similar or close to that obtained from the single instance above. In particular, the underlying estimation procedure provides reliable standard errors for the non-linear utility functions employed in this paper. We conclude that estimations based on a single simulation instance are sufficient for further analyses.

**Table 2:**
Mean and standard error of coefficients obtained by simulation.

| Variable | Synthetic | Linear  | HTF     | STF1    | STF2    | Power   |
|----------|-----------|---------|---------|---------|---------|---------|
| Cost     | -0.600    | -0.632  | -0.602  | -0.605  | -0.605  | -0.608  |
|          |           | (0.020) | (0.020) | (0.020) | (0.020) | (0.020) |
|          |           | [0.019] | [0.019] | [0.019] | [0.019] | [0.020] |
| Time     | -0.100    | -0.077  | -0.100  | -0.105  | -0.109  | -0.018  |
|          |           | (0.004) | (0.008) | (0.011) | (0.013) | (0.008) |
|          |           | [0.004] | [0.008] | [0.014] | [0.017] | [0.007] |
| Alpha    | 5.000     | ---     | 4.901   | 5.433   | 6.326   | 1.505   |
|          |           |         | (1.112) | (1.454) | (1.886) | (0.147) |
|          |           |         | [1.020] | [1.770] | [2.340] | [0.175] |

(.) Standard error obtained from 500 simulation instances.
[.] Standard error of the corresponding estimated parameter in Table 1.

## 4. The value of travel time savings

The value of travel time savings (VTTS) is calculated as the compensatory variation per unit of travel time. The compensatory variation is the maximum amount of money a person is willing to pay for time savings. This payment keeps the person on the same utility level as in the situation without the time savings. In this context the compensatory variation for specific time savings is defined as the corresponding cost increase, which in turn is equivalent to an income reduction. Hence, the VTTS is defined by ( 6 ). The resulting formulas for the various transformations are given by ( 7 ) – ( 11 ). Costs are measured in Swiss francs (CHF) and time in minutes. The value of travel time savings for finite amounts $\Delta T$ of time differences is expressed in CHF per hour.[12]

$$\text{VTTS} = -\frac{\Delta C}{\Delta T}\bigg|_{\Delta V=0} * 60 \qquad (6)$$

$$\text{VTTS}_{\text{Linear}} = \frac{\beta_T}{\beta_C} * 60 \qquad (7)$$

$$\text{VTTS}_{\text{HTF}} = \begin{cases} 0 & \text{abs}(\Delta T) < \alpha_{HTF} \\ \beta_T/\beta_C * \left(1 - \frac{\alpha_{HTF}}{\text{abs}(\Delta T)}\right) * 60 & \text{abs}(\Delta T) \geq \alpha_{HTF} \end{cases} \qquad (8)$$

$$\text{VTTS}_{\text{STF1}} = \frac{\beta_T}{\beta_C} * \left(1 - \alpha_{STF1} \tanh\left(\frac{\Delta T}{\alpha_{STF1}}\right) * \Delta T^{-1}\right) * 60 \qquad (9)$$

$$\text{VTTS}_{\text{STF2}} = \frac{\beta_T}{\beta_C} * \left(1 - 1/\sqrt{\left(\frac{\Delta T}{\alpha_{STF2}}\right)^2 + 1}\right) * 60 \qquad (10)$$

---
[12] For this reason each VTTS formula is multiplied by 60 [min/hr].

$$\text{VTTS}_{\text{Power}} = \frac{\beta_T}{\beta_C} * sign(\Delta T) * abs(\Delta T)^{\alpha_{Power}} * \Delta T^{-1} * 60 \qquad (11)$$

Figure 3 depicts the corresponding VTTS for the synthetic data. As expected, the models considering thresholds exhibit a lower VTTS for smaller time changes and a higher VTTS for larger time changes in comparison to the linear model.

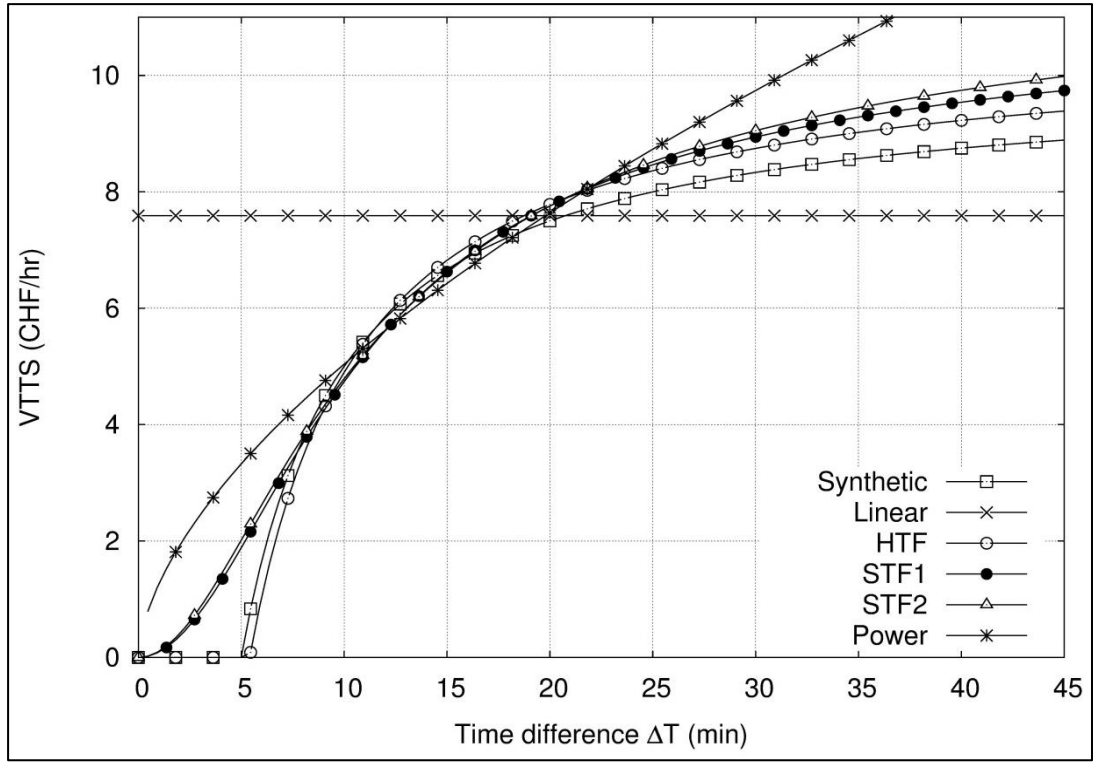

**Figure 3:** Value of travel time savings (VTTS) for the indicated transformation functions based on the synthetic data.

Furthermore, an asymptotic VTTS was calculated for the two STF and the HTF. This is the VTTS for $\Delta T \rightarrow \pm\infty$, which is in this case simply the ratio of time and cost coefficients. The ratios of time and cost coefficients for the different specifications are reported in Table 1. However, an asymptotic VTTS does not exist for the power function. Calculating the VTTS as the ratio of the time and cost coefficient would result in an incorrect value of 1.30 CHF/hour. Thus, the power function specification appears to be problematic for estimating the correct value of time, if thresholds exist.[13] It is also easy to see that the linear specification shows the highest deviation from the synthetic data. Thus, the problems of the linear and power function in estimating the correct time coefficient affect the VTTS calculations as well.

---

[13] This is not necessarily the case if people do really exhibit a sensitivity of infinity for large time changes, which, however, seems to be unrealistic to us.

In addition, we tested synthetically generated data with several other threshold levels. The results with respect to the asymptotic VTTS are depicted in Figure 4.[14] To visualise the fact that the linear specification systematically underestimates the asymptotic VTTS if a hard threshold is present in the data, the 95% confidence interval for each VTTS point estimate is also shown in the figure. These intervals have been determined according to the multivariate normal simulation method described for instance by Armstrong et al. (2001) or Bliemer and Rose (2013). We based our calculations on 100 000 draws from the bivariate normal distribution. Furthermore, we were able to validate our results with the asymptotic t-test method described in Armstrong et al. (2001).[15]

Figure 4 shows that the synthetic asymptotic VTTS of 10 CHF/hour lies within the confidence intervals of HTF, STF1, and STF2 but outside the interval of the linear function. Furthermore, the plot for the linear function reveals that the underestimation of the true VTTS becomes more severe with an increasing time threshold.[16] For STF1 and STF2 the mean deviates stronger from the synthetic reference the larger the time threshold; the corresponding confidence intervals also become wider. However, this stems from the fact that data is generated according to the hard threshold function. Nevertheless, considering absolute deviations, the mean asymptotic VTTS of all threshold transformations are closer to the synthetic one than the linear function results.

---

[14] The estimation results are qualitatively equal to these in Table 1 and therefore not presented here.

[15] As Armstrong et al. (2001) we obtain similar results with the t-test and multivariate normal simulation method. According to Armstrong et al. (2001) the multivariate normal simulation method is one of most accurate methods.

[16] For synthetic data with a threshold of just one minute, an insignificant threshold parameter has been obtained in all specifications. However, based on this information, one should not conclude that thresholds are detectable only from two minutes upwards. Rather, this depends strongly on the distribution of the time differences and especially on the number of observation with time differences close to the threshold.

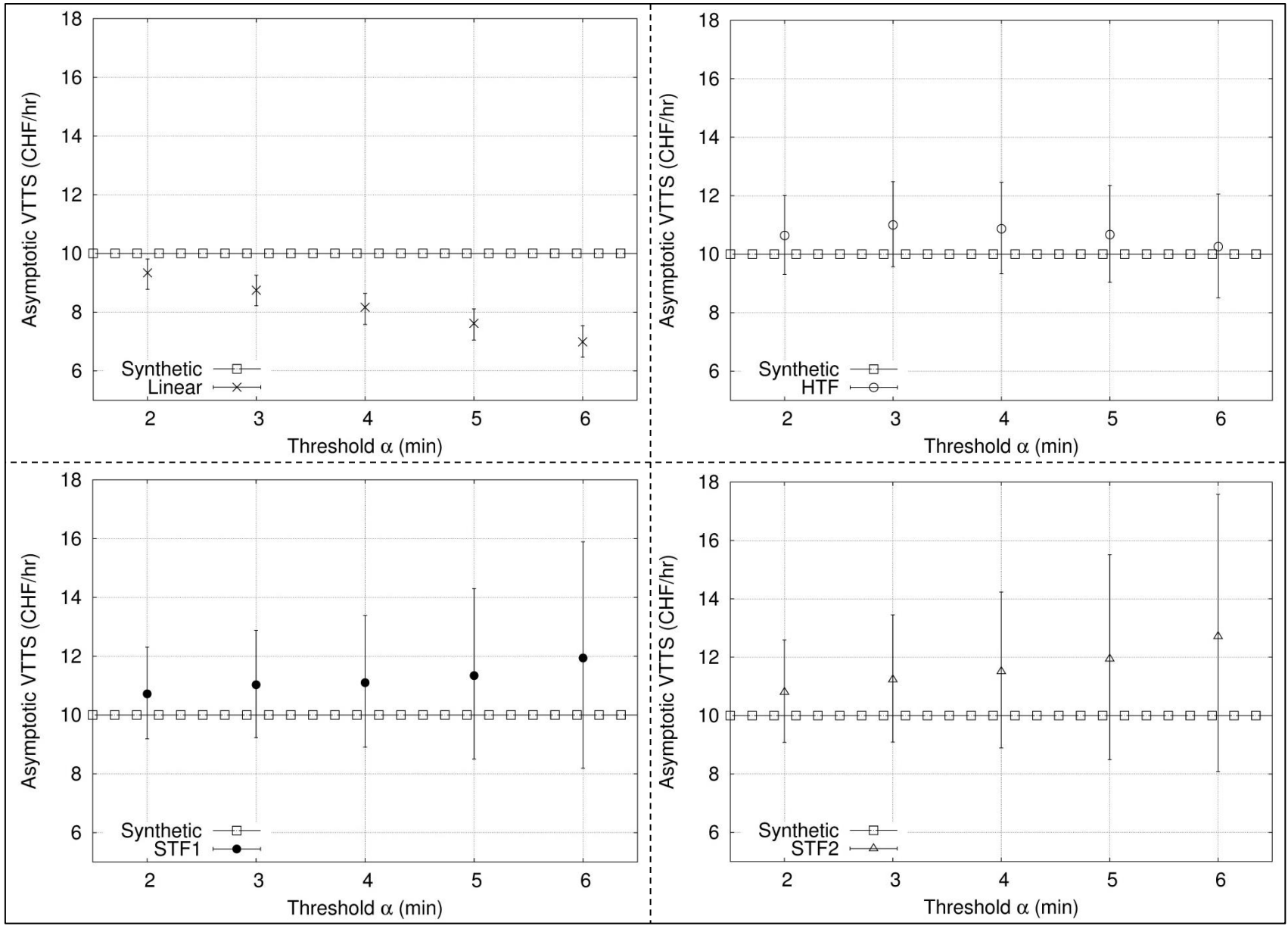

**Figure 4:** Asymptotic value of travel time savings (VTTS) with 95% confidence intervals for the indicated transformation functions and threshold sizes. The synthetic data has been generated with a hard threshold and the respective threshold size.

With regard to transport project appraisal, two major arguments against a lower valuation of small travel time savings (as shown above) can be found in the literature. First, on the level of individuals, it might be questionable whether thresholds really exist, even if they can be found in stated choice data (e.g. Fowkes, 1999; Tsolakis et al., 2011). Individuals might show these thresholds in choice experiments, but this would be more or less an artificial result of the survey method. A possible explanation for this is that people do not consider the opportunity of rescheduling their activities in the short run to make use of small time savings (Mackie et al. 2001). Thus, the central question is whether transport users' valuations differ with the size of the savings *under real-world conditions*. In this regard, an analysis based on revealed preference data might help. However, this approach raises problems of misperception, as trip makers simply might not perceive small time differences and consequently do not consider them in their choice decision. Thus, valuation and perception thresholds may be confounded when considering real choices.[17] The perception effect might be relevant for modelling and forecasting route choices in reality, but the matter of interest for a benefit-cost analysis seems to be whether or not transport users care about them, not whether or not they perceive them. The latter issue seems to be just a problem of incomplete information. With stated choice data, the perception effect can be ruled out.

Second, time savings might add up across different (transport) projects (Fowkes, 1999; Mackie et al., 2001). Therefore, all transport users exceed (after several projects) the threshold.[18] Even if people do not add up explicitly in their minds, it can be argued that it is just important that all time differences are actually *received* by the individuals. However, when combining time savings from different activities, people probably must reschedule their activities (Tsolakis et al., 2011, p. 14). In circumstances in which individual timetables underlie certain restrictions and, therefore, rescheduling is not possible, a threshold might still be of relevance.[19] Furthermore, we see another counterargument related to the cost of thinking. People might avoid the cognitive costs of rescheduling their activities when time savings are sufficiently small.

Closely related to this, it has been argued (Fowkes, 1999) that when discounting small time savings, the breaking up of a large project into smaller parts would result in different travel time savings benefits. The inclusion of thresholds would therefore favour large-scale projects and penalize small-scale ones, since the threshold will only be exceeded with large scaled projects. However, one should not ignore that a large scale project might have a higher total effect (e.g. on the modal split) than several smaller ones with the same accumulated time savings if thresholds are present and people continuously update their reference points.

---

[17] This distinction between perception and valuation is also mentioned by Tsolakis et al. (2011, p. 14).

[18] The argument is more sophisticated and has already been formalised by Fowkes and Wardman (1988). The additivity of time savings is the key element of it. For an in-depth discussion see Fowkes (1999).

[19] A similar argument can be found in Tsolakis et al. (2011, p. 12).

With respect to the two arguments presented above, we should point out that we argue neither for nor against a discounted unit value for small travel time savings in transport project appraisal. An in depth analysis of these questions would go beyond the scope of this paper. Nevertheless, the concerns raised so far against the discounted unit value approach do not imply that thresholds can be completely ignored in model estimation. We want to emphasise that, irrespective of whether one agrees with above arguments or not, the inclusion of thresholds in the modelling approach seems to be necessary. Otherwise, as we have shown, the estimated VTTS can be substantially biased downwards. This is because even an *artificial* threshold that people just show in a stated choice framework but not in real-life situations influences the valuation of attributes in the experiment. In this sense, the transformation function corrects for the threshold, which might have been just caused by the survey method.[20] Thus, even if the discounted unit value approach for transport scheme appraisal is rejected, the asymptotic VTTS based on a threshold model should be used for project assessment if a time threshold exists.

## 5. Application to stated choice data

The data we use originate from route choice experiments for commuting trips by train in Switzerland. In these experiments, respondents had to choose between two routes which were characterised by the attributes travel time, travel cost, headway ($H$) and the number of changes ($K$). The data contain around 1 600 observations from roughly 180 respondents. The average travel time in the data is 30 minutes and the mean of the cost variable is 12 CHF. The range of time differences varies from one minute to around 45 minutes with 20 per cent of the observations less than or equal to two minutes.[21] A binary logit model with the following deterministic utility function has been estimated.[22]

$$\Delta V(\Delta T, \Delta C, \Delta H, \Delta K, I, T) = \beta_T * f_r(\Delta T, \alpha_r) + \beta_C \Delta C * \left(\frac{I}{\bar{I}}\right)^{\lambda_I} * \left(\frac{T}{\bar{T}}\right)^{\lambda_T} + \beta_H \Delta H + \beta_K \Delta K \quad (12)$$

The above formulation of utility additionally considers the elasticities $\lambda_I$ and $\lambda_T$. Negative parameter values imply an increase of the value of time with income ($I$) and the average travel time of the two offered alternatives ($T$), respectively. Both variables are normalised to their average. The values of $\lambda_I$ and $\lambda_T$ have to be estimated. In contrast to Axhausen et al. (2008) and others, we did not estimate the elasticity for distance because people usually consider travel time rather than distance when deciding on a trip by train. This has been confirmed by preliminary tests, which showed a clear improvement in model fit when using elasticity of time instead of distance. The estimation results for the different transformations mentioned above are summarised in Table 3.

---

[20] Note that Börjesson and Eliasson (2014, p. 153) also state that it is important to control for size effects in value of time studies, irrespective of whether they are considered for project assessment or not.

[21] For a more detailed description of the complete database and the survey design see Axhausen et al. (2008). Based on preliminary tests we restricted our analysis to rail route choices and, therefore, we use just a part of the whole database.

[22] This function is based on the model of Axhausen et al. (2008).

**Table 3:**
Estimation results for stated choice data.

| Variable | Linear | HTF | STF1 | STF2 |
|---|---|---|---|---|
| Cost | -0.305 * | -0.274 * | -0.285 * | -0.286 * |
| Time | -0.127 * | -0.159 * | -0.151 * | -0.152 * |
| Alpha | --- | 2.760 * | 2.170 # | 2.310 # |
| Headway | -0.050 * | -0.051 * | -0.051 * | -0.051 * |
| Changes | -1.420 * | -1.430 * | -1.430 * | -1.430 * |
| Income Elasticity | -0.252 * | -0.251 * | -0.250 * | -0.249 * |
| Time Elasticity | -0.489 * | -0.347 * | -0.387 * | -0.391 * |
| Scale [a] | 0.797 [*] | 0.787 [*] | 0.790 [*] | 0.790 [*] |
| VTTS [b] | 24.98 | 34.82 | 31.79 | 31.89 |
| Null-LL | | -1110.422 | | |
| Final-LL | -687.064 | -684.184 | -684.651 | -684.798 |
| LR test against Linear | --- | 0.02 | 0.03 | 0.03 |

\*, #, + Significant on 1%, 5%, and 10% levels, respectively.
[.] Significance level for null hypothesis that parameter is equal to one.
[a] Controls for error scale differences.
[b] Asymptotic value of travel time savings in CHF per hour for mean income and travel time.

The scale variable allows for different magnitudes in the error components of different user groups, in our case car drivers and rail users, who both participated in the experiments.[23] The scale parameter of the rail users has been set to unity. An estimated scale of less than one for the group of car drivers indicates that they exhibit a higher error variance than rail users.[24] Hence, the variance of the unobserved factors is greater for car than for rail users.

The linear model is always a special case of the threshold models. Likelihood-ratio tests show that the HTF and the STF are significantly better than the linear model on a 2 and 3 per cent significance level, respectively. Interestingly, as with the synthetic data, we observe a difference in the log-likelihood value of around 1 per 500 observations. With a test developed by Horowitz (1983) to compare non-nested models, HTF and STF can be tested against each other. The test revealed that the STF formulations perform not significantly worse than the HTF model.[25] In addition, the power function formulation proved to be significantly worse than the other three models, and was therefore omitted from the analysis. Furthermore, the power coefficient was not significantly different from one. This is a further indication of the problems related to the power function for detecting thresholds correctly.

---

[23] See Train (2009), sections 2.5.2 and 3.2, for an in depth discussion of the error scale.

[24] This is in line with the findings of König et al. (2004).

[25] We based our calculations on formula 52 in Horowitz (1983, p. 336). Note that a different formulation can be found in the literature, e.g. in Ben-Akiva and Lerman (2007, p. 172). We however, regard the original formula of Horowitz as the correct one.

Across all three threshold specifications, significant threshold parameters have been estimated. They indicate a threshold between two and three minutes. All other coefficients are significant as well and within the range of expectations. Figure 5 depicts the utility difference against the time difference and reveals great similarities between the two STF.[26]

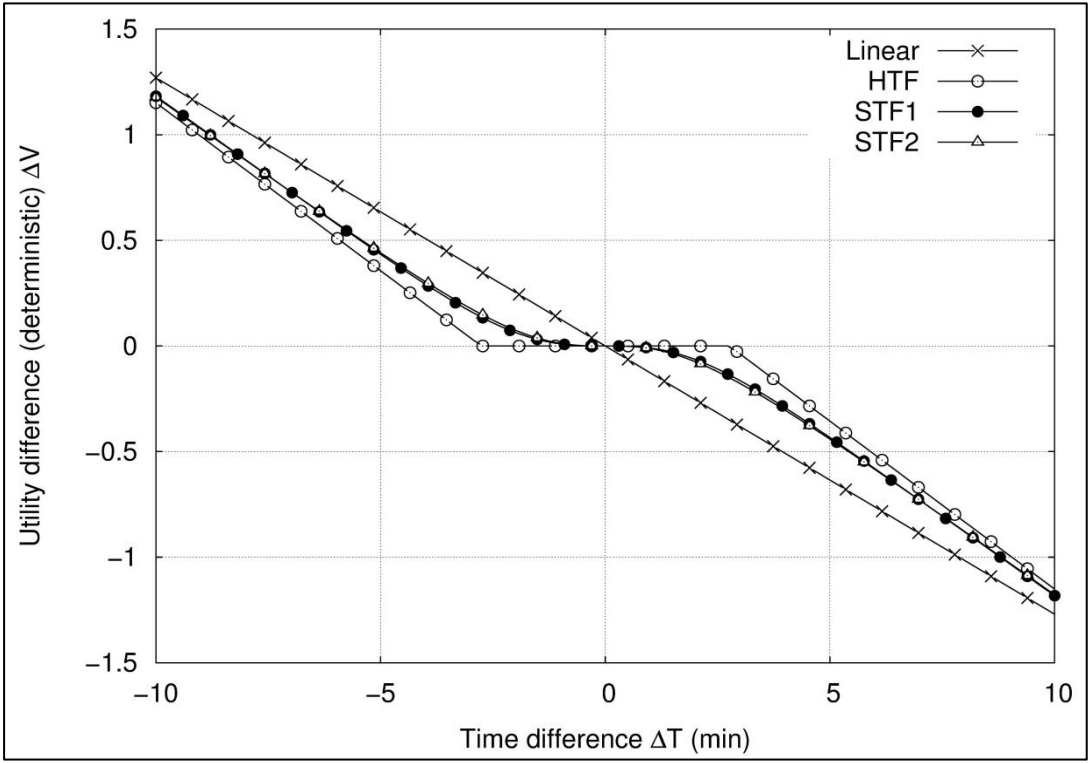

**Figure 5:** Estimated utility functions for the indicated transformation functions based on stated choice data with 1 600 observations.

Finally, in Figure 6 the resulting VTTS are plotted for mean income and travel time. As with the synthetic data, the threshold formulations show a lower VTTS for smaller time changes and a higher VTTS for larger time changes in comparison to the linear model. The asymptotic values are reported in Table 3. Again, as with the synthetic data, we can observe that the inclusion of thresholds leads to substantially higher asymptotic VTTS values.

---

[26] All other explanatory variables have been set to zero to reduce dimensionality.

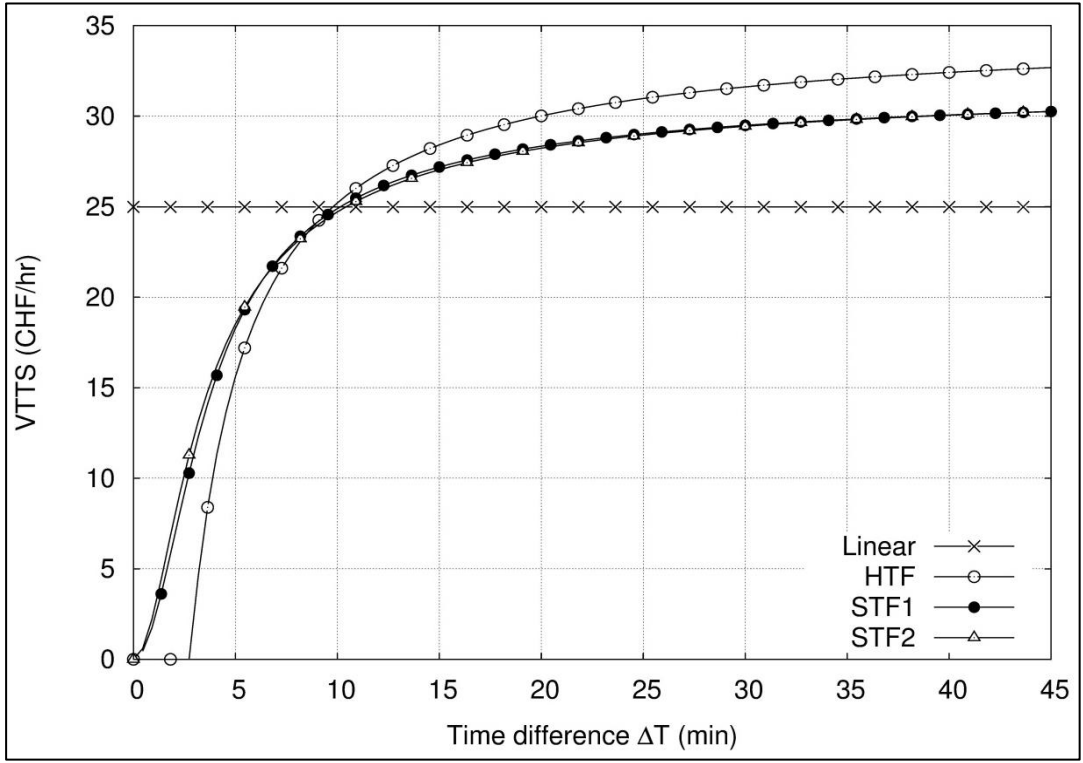

**Figure 6:** Value of travel time savings (VTTS) for the indicated transformation functions for stated choice data.

The two soft threshold functions presented in this paper constitute smooth approximations to the hard threshold function commonly used to detect thresholds. A noteworthy feature of these functions is that a time difference well above the threshold is transformed as well, e.g. with a threshold of five minutes, a fifteen minute time difference is perceived as ten minutes. One might argue that this seems inappropriate if, for example, re-scheduling impossibilities drive the occurrence of thresholds. A transformation function which reverts to linearity for large time differences, as Figure 7 depicts, can overcome this objection. We therefore tested the reverting transformation function

$$f_{Reverting}(\Delta T, \alpha_{Reverting}) = \frac{\Delta T}{1 + e^{\alpha_{Reverting} - abs(\Delta T)}} \qquad (13)$$

for the stated choice data but could not determine a significant improvement. To analyse this issue profoundly will be the topic of future research. In summary, we find that detecting a threshold as such is more relevant than determining the specific form of threshold.

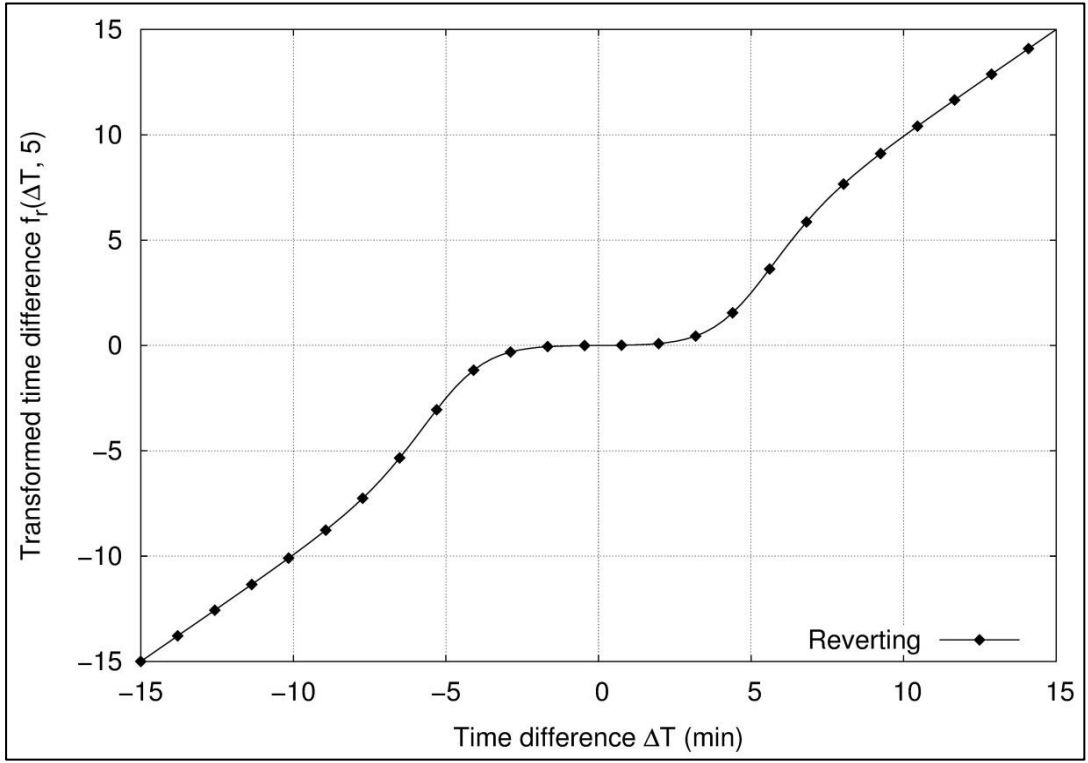

**Figure 7:** Generic form of the transformation function ( 13 ); for illustration purposes this function is plotted for a threshold parameter $\alpha_r = 5$ (min).

## 6. Conclusion

In this contribution, we tested for thresholds in individual choice behaviour. We focused on empirical issues in estimating such thresholds with discrete choice models and the consequence of ignoring them in model estimation. From our point of view, this issue has to be addressed separately from the question whether thresholds should be considered in benefit-cost analyses. We proposed different functions, including a piecewise linear function with a hard threshold, two smooth functions with soft thresholds, and a power function. To the best of our knowledge, the two soft threshold functions have not been used elsewhere in the literature. We applied these functions to synthetic and stated choice data. Estimating the generated data showed that many observations are necessary even to detect hard thresholds. Interestingly, for both the synthetic and the real data, we observed a difference in the log-likelihood value of around 1 per 500 observations between the linear and the threshold models. As stated choice data we employed the rail route choice experiments from the Swiss value of travel time study (Axhausen et al., 2008). The results indicate a time threshold between two and three minutes. However, the cause of this threshold cannot be inferred from the data. We were also not able to detect whether it is a hard or soft one. Additionally, we found that for this data it does not matter whether the transformation function reverts to linearity or not. Furthermore, tests with the synthetic data showed that all functions except the power transformation worked reasonably well in reproducing the known values. It has to be emphasised that the performance of the different transformations depends on the underlying data generation process. Nevertheless, we think that it is more plausible that trip makers' sensitivities converge to a limit instead of rising to infinity.

The detection of thresholds affects the inferred value of time. According to the estimates, small travel time savings should be valued at a lower rate than larger ones. Moreover, the large ones should be valued even higher than currently. Arguments against such a treatment in project scheme appraisal have been raised in the literature before. However, as we have shown in this paper, thresholds should not be ignored in model estimation, even if they are just an artificial result of the survey method. Otherwise, the estimated VTTS may be substantially biased downwards. Thus, even if the discounted unit value approach for transport scheme appraisal is rejected, the asymptotic VTTS based on a threshold model should be considered for project assessment if a time threshold exists. Furthermore, regardless the monetisation issues discussed in this paper, thresholds might be important for predicting choice behaviour.

Clearly, there remain numerous unresolved questions regarding the value of time for project evaluation. In closing, we emphasise that the estimation procedure and transformation functions presented in this paper may be useful for modelling more general choice situations beyond route choice, as well as for modelling a smooth version of indifference (utility) thresholds considered by Cantillo et al. (2010).

## Acknowledgements

We wish to thank K. W. Axhausen (ETH Zurich) for providing us with the stated choice data and three anonymous reviewers for their valuable comments. Financial support for this research was provided by the German Research Foundation (DFG).

## References

Armstrong, P., Garrido, R., Ortúzar, J.d.D., 2001. Confidence intervals to bound the value of time. Transportation Research Part E: Logistics and Transportation Review. 37(2-3), 143-161.

Axhausen, K.W., Hess, S., König, A., Abay, G., Bates, J.J., Bierlaire, M., 2008. Income and distance elasticities of values of travel time savings: New Swiss results. Transport Policy 15 (3), 173-185.

Bates, J., Whelan, G., 2001. Size and sign of time savings. Institute of Transport Studies, University of Leeds. Working Paper 561.

Ben-Akiva, M., Lerman, S.R., 2007. Discrete Choice Analysis: Theory and Application to Travel Demand. MIT Press, Cambridge/London.

Bliemer, M.C.J., Rose, J.M., 2013. Confidence Intervals of Willingness-to-Pay for Random Coefficient Logit Models. Transportation Research: Part B: Methodological. 58, 199-214.

Börjesson, M., Eliasson, J., 2014. Experiences from the Swedish value of time study. Transportation Research Part A: Policy and Practice 59, 144-158.

Cantillo, V., Amaya, J., Ortúzar, J.d.D., 2010. Thresholds and indifference in stated choice surveys. Transportation Research Part B: Methodological 44(6), 753-763.

Cantillo, V., Heydecker, B., Ortúzar, J.d.D., 2006. A discrete choice model incorporating thresholds for perception in attribute values. Transportation Research Part B: Methodological 40(9), 807-825.

Fosgerau, M., 2007. Using nonparametrics to specify a model to measure the value of travel time. Transportation Research Part A: Policy and Practice 41(9), 842-856.

Fosgerau, M., Jensen, T.L., 2003. Economic appraisal methodology-controversial issues and Danish choices. Proceedings of the European Transport Conference (ETC) 2003, 8-10 October 2003, Strasbourg, France.

Fowkes, A.S., Wardman, M., 1988. The design of stated preference travel choice experiments - with special reference to interpersonal taste variations. Journal of Transport Economics and Policy 22(1), 27-44.

Fowkes, A.S., 1999. Issues in evaluation: a justification for awarding all time savings and losses, both small and large, equal unit values in scheme evaluation. in: Accent/Hague Consulting Group (Ed.), The Value of Travel Time on UK Roads - Report to DETR, 341-359.

Givoni, M., 2008. A Comment on 'The Myth of Travel Time Saving'. Transport Reviews. 28(6), 685-688.

Gunn, H., 2001. Spatial and temporal transferability of relationships between travel demand, trip cost and travel time. Transportation Research Part E: Logistics and Transportation Review 37(2-3), 163-189.

Hensher, D. A. ,2001. Measurement of the valuation of travel time savings. Journal of Transport Economics and Policy 35(1), 71-98.

Hjorth, K., Fosgerau, M., 2012. Using prospect theory to investigate the low marginal value of travel time for small time changes. Transportation Research Part B: Methodological 46(8), 917-932.

Horowitz, J.L., 1983. Statistical comparison of non-nested probabilistic discrete choice models. Transportation Science 17(3), 319-350.

Hultkrantz, L., Mortazavi, R., 2001. Anomalies in the value of travel-time changes. Journal of Transport Economics and Policy 35(2), 285-300.

Ironmonger, D., Norman, P., 2008. Improvements in Transport Infrastructure are designed to Increase Travel Speed: Comments on 'The Myth of Travel Time Saving'. Transport Reviews. 28(6), 694-698.

König, A., Axhausen, K.W., Abay, G., 2004. Zeitkostenansätze im Personenverkehr. Final report for SVI 2001/534, Schriftenreihe, 1065, Bundesamt für Strassen, UVEK, Bern.

Krishnan, K.S., 1977. Incorporating thresholds of indifference in probabilistic choice models. Management Science 23(11), 1224-1233.

Li, C., Hultkrantz, L., 2004. A stochastic threshold model for estimating the value of travel time. in: B. Mao, Z. Tian, Q. Sun (Ed.), Traffic and Transportation Studies, Proceedings of ICTTS 2004. Science Press, Beijing.363-373.

Lioukas, S.K., 1984. Thresholds and transitivity in stochastic consumer choice: a multinomial logit analysis. Management Science 30(1), 110-122.

Lyons, G., 2008. A Comment on 'The Myth of Travel Time Saving'. Transport Reviews. 28(6), 706-709.

Mackie, P., 2008. Who Knows Where the Time Goes? A Response to David Metz. Transport Reviews. 28(6), 692-694.

Mackie, P.J., Jara-Diaz, S.R., Fowkes, A.S., 2001. The value of travel time savings in evaluation. Transportation Research Part E: Logistics and Transportation Review 37(2-3), 91-106.

Mackie, P.J., Wardman, M., Fowkes, A.S., Whelan, G., Nellthorp, J., 2003. Values of travel time savings in the UK. Report to Department for Transport.

Metz, D., 2008a. The myth of travel time saving. Transport Reviews 28(3), 321-336.

Metz, D., 2008b. Response to the Responses. Transport Reviews. 28(6), 713-715.

Noland, R.B., 2008. Understanding Accessibility and Road Capacity Changes: A Response in Support of Metz. Transport Reviews. 28(6), 698-706.

Schwanen, T., 2008. Reflections on Travel Time Savings: Comments to David Metz. Transport Reviews. 28(6), 709-713.

Small, K.A., 2012. Valuation of travel time. Economics of Transportation. 1(1–2), 2-14.

Small, K.A., Verhoef, E., 2007. The economics of urban transportation. Routledge, London/New York.

Stathopoulos, A., Hess, S., 2012. Revisiting reference point formation, gains-losses asymmetry and non-linear sensitivities with an emphasis on attribute specific treatment. Transportation Research: Part A: Policy and Practice 46(10), 1673-1689.

Train, K.E., 2009. Discrete choice methods with simulation. Cambridge University Press, Cambridge.

Tsolakis, D., Shackleton, J., Makwasha, T., 2011. Small travel time savings: treatment in project evaluations. Austroads Publication No. AP–R392-11.

Van Wee, B., Rietveld, P., 2008. 'The Myth of Travel Time Saving': A Comment. Transport Reviews. 28(6), 688-692.

Welch, M., Williams, H., 1997. The sensitivity of transport investment benefits to the evaluation of small travel-time savings. Journal of Transport Economics and Policy 31(3), 231-254.